\documentclass[letter, longauth, traditabstract]{aa} 
\usepackage{graphicx}
\usepackage{txfonts}
\usepackage{natbib}

\newcommand\msol{\mbox{$M_\odot$}}
\newcommand\lsol{\mbox{$L_\odot$}}

\newcommand\asec{\mbox{$^{\prime\prime}$}}%
\newcommand\amin{\mbox{$^\prime$}}%
\begin{document}
   \title{The \emph{Herschel}\thanks{ {\it Herschel} is an ESA space observatory with science instruments provided by European-led Principal Investigator consortia and with important participation from NASA.} first look at protostars in the Aquila Rift\thanks{
Figures~\ref{fig:pacs_images}--\ref{fig:distribution} are only available in electronic form at
http://www.aanda.org.}}
   \titlerunning{{\it Herschel} first look on protostars in the Aquila Rift}

   \author{S. Bontemps\inst{1,2,3}
          \and
            Ph. Andr\'e\inst{1}
          \and
            V. K\"onyves\inst{1}
          \and
            A. Men'shchikov\inst{1}
          \and
            N. Schneider\inst{1}
          \and
           A. Maury\inst{4}
          \and
           N. Peretto\inst{1}
          \and
          D. Arzoumanian\inst{1}
          \and
          M. Attard\inst{1}
          \and
            F. Motte\inst{1}
          \and
            V. Minier\inst{1}
          \and
            P. Didelon\inst{1}
          \and
            P. Saraceno\inst{5}
          \and
            A. Abergel\inst{6}
          \and
            J.-P. Baluteau\inst{7}
          \and
            J.-Ph. Bernard\inst{8}
          \and
            L. Cambr\'esy\inst{9}
          \and
            P. Cox\inst{10}
          \and
            J. Di Francesco\inst{11}
          \and
            A. M. Di Giorgo\inst{5}
          \and
            M. Griffin\inst{12}
          \and
            P. Hargrave\inst{12}
          \and
            M. Huang\inst{13}
          \and
            J. Kirk\inst{12}
          \and
            J. Li\inst{13}
          \and
            P. Martin\inst{14}
          \and
            B. Mer{\'{\i}}n\inst{15}
          \and
            S. Molinari\inst{5}
          \and
            G. Olofsson\inst{16}
          \and
            S. Pezzuto\inst{5}
          \and
            T. Prusti\inst{15}
          \and
            H. Roussel\inst{17}
          \and
            D. Russeil\inst{7}
          \and
            M. Sauvage\inst{1}
          \and
            B. Sibthorpe\inst{18}
          \and
            L. Spinoglio\inst{5}
          \and
            L. Testi\inst{4, 19}
          \and
            R. Vavrek\inst{15}
         \and
            D. Ward-Thompson\inst{12}
          \and
            G. White\inst{20,21}
          \and
            C. Wilson\inst{22}
          \and
            A. Woodcraft\inst{23}
          \and
            A. Zavagno\inst{7}
          }

   \institute{Laboratoire AIM, CEA/DSM--CNRS--Universit\'e Paris Diderot, IRFU/Service d'Astrophysique, C.E. Saclay,
              Orme des Merisiers, 91191 Gif-sur-Yvette, France
        \and
             CNRS/INSU, Laboratoire d'Astrophysique de Bordeaux, UMR 5804, BP 89, 33271 Floirac cedex, France
         \and     
              Universit\'e de Bordeaux, OASU, Bordeaux, France
         \and
             European Southern Observatory,  Karl Schwarzschild Str. 2, 85748 Garching, Germany
         \and
             INAF-IFSI, Fosso del Cavaliere 100, 00133 Roma, Italy
         \and
             IAS, CNRS-INSU--Universit\'e Paris-Sud, 91435 Orsay, France
         \and
             Laboratoire d'Astrophysique de Marseille, CNRS/INSU--Universit\'e de Provence, 13388
             Marseille cedex 13, France
         \and
             CESR \& UMR 5187 du CNRS/Universit\'e de Toulouse, BP 4346, 31028 Toulouse Cedex 4, France
         \and          
             Observatoire astronomique de Strasbourg, UMR 7550 CNRS/Universit\'e de Strasbourg, 11 rue de l'Universit\'e, 67000, Strasbourg
         \and          
             IRAM, 300 rue de la Piscine, Domaine Universitaire, 38406 Saint Martin d'H\`eres, France        
         \and         
             National Research Council of Canada, Herzberg Institute of Astrophysics,
             University of Victoria, Department of Physics and Astronomy, Victoria, Canada
         \and          
             School of Physics and Astronomy, Cardiff University, Queens Buildings The Parade, Cardiff CF24 3AA, UK
         \and          
             National Astronomical Observatories, Chinese Academy of Sciences, Beijing 100012, China
         \and
             CITA \& Dep. of Astronomy and Astrophysics, University Toronto, Toronto, Canada
         \and
             Herschel Science Center, ESAC, ESA, PO Box 78, Villanueva de la Ca\~nada, 28691 Madrid, Spain
         \and
             Department of Astronomy, Stockholm Observatoty, AlbaNova University Center, Roslagstullsbacken 21, 10691 Stockholm, Sweden
         \and
             Institut d'Astrophysique de Paris, UMR7095 CNRS, UniversitŽ Pierre et Marie Curie, 98 bis Boulevard Arago, 75014 Paris, France
         \and
             Astronomy Technology Centre, Royal Observatory Edinburgh, Blackford Hill, EH9 3HJ, UK
         \and
             INAF--Osservatorio Astrofisico di Arcetri, Largo Fermi 5, 50125 Firenze, Italy
         \and
             Science and Technology Facilities Council, Rutherford Appleton Laboratory, Chilton, Didcot OX11 0NL, UK             
         \and
             Department of Physics \& Astronomy, The Open University, Walton Hall, Milton Keynes MK7 6AA, UK
         \and
             Department of Physics and Astronomy, McMaster University, Hamilton, ON L8S 4M1, Canada
         \and
             SUPA, Institute for Astronomy, Edinburgh University, Blackford Hill, Edinburgh EH9 3HJ, UK
          }

   \date{Received ; accepted }

 
  \abstract
  { As part of the science demonstration phase of the {\it Herschel} mission of the Gould Belt Key Program, the Aquila Rift molecular complex has been observed. The complete $\sim 3.3^\circ \times 3.3^\circ$ imaging with SPIRE 250/350/500$\,\mu$m and PACS 70/160$\,\mu$m allows a deep investigation of embedded protostellar phases, probing of the dust emission from warm inner regions at 70 and 160$\,\mu$m to the bulk of the cold envelopes between 250 and 500$\,\mu$m.
    We used a systematic detection technique operating simultaneously on all  {\it Herschel} bands to build a sample of protostars. Spectral energy distributions are derived to measure luminosities and envelope masses, and to place the protostars in an $M_{\rm env} - L_{\rm bol}$ evolutionary diagram. The spatial distribution of protostars indicates three star-forming sites in Aquila, with W40/Sh2-64 H{\tiny II} region by far the richest. Most of the detected protostars are newly discovered. For a reduced area around the Serpens South cluster, we could compare the  {\it Herschel} census of protostars with {\it Spitzer} results. The {\it Herschel} protostars are   younger than in {\it Spitzer} with 7 Class~0 YSOs newly revealed by {\it Herschel}. For the entire Aquila field, we find a total of $\sim45-60$ Class 0 YSOs discovered by {\it Herschel}. This confirms the global statistics of several hundred Class~0 YSOs that should be found in the whole  Gould Belt survey.  }


   \maketitle
%

\section{Introduction}
\label{sect:intro}


During the main accretion phase, protostars are deeply embedded in their collapsing envelopes and parent clouds. They are so embedded that they radiate mostly at long wavelengths, making their detection and study difficult from the ground (e.g. \citealp{andre2000, difrancesco2007}). Protostars, or young stellar objects (YSOs), in the solar neighborhood have been extensively surveyed, but a complete and unbiased census of all protostars in nearby molecular clouds is lacking.  The census of embedded YSOs provided by  IRAS and near-IR studies in the 1980s and 1990s was far from complete even in the nearest clouds. Thanks to their high sensitivity and good spatial resolution in the mid-infrared,  ISO and, more recently,  {\it Spitzer} could perform more complete surveys in all major nearby star-forming regions (e.g. \citealp{nordh1996, bontemps2001, kaas2004, allen2007, evans2009}). The population of the youngest protostars, the Class~0 YSOs, can however not be properly surveyed solely in the near and mid-infrared. These youngest objects remain weak or undetected shortward of $\sim20\,\mu$m. 


The {\it Herschel} Gould Belt Survey \citep{andre2010} is a key program of the ESA Herschel Space Observatory \citep{pilbratt2010}. It employs the SPIRE \citep{griffin2010} and PACS \citep{poglitsch2010} instruments to do photometry in large-scale far-infrared images at an unprecedented spatial resolution and sensitivity. The Aquila Rift region has been chosen to be observed for the science demonstration phase of  {\it Herschel} for this survey. 

Our 250/350/500~$\mu$m SPIRE and 70/160~$\mu$m PACS images of the Gould Belt provide the first access to the critical spectral range of the far-infrared to submillimeter regimes to cover the peak of the spectral energy distributions (SEDs) of the cold phase of star formation at a high enough spatial resolution to separate individual objects. The {\it Herschel} surveys therefore allow an unprecedented, unbiased census of starless cores \citep{konyves2010}, embedded protostars (this work), and cloud structure \citep{menshchikov2010}, down to the lowest column densities \citep{andre2010}.
This survey yields the first accurate far-infrared photometry, hence good luminosity and mass estimates, for a comprehensive view of all early evolutionary stages.


   \begin{figure}[hbtp]
  \centering
  \includegraphics[angle=270,width=9cm]{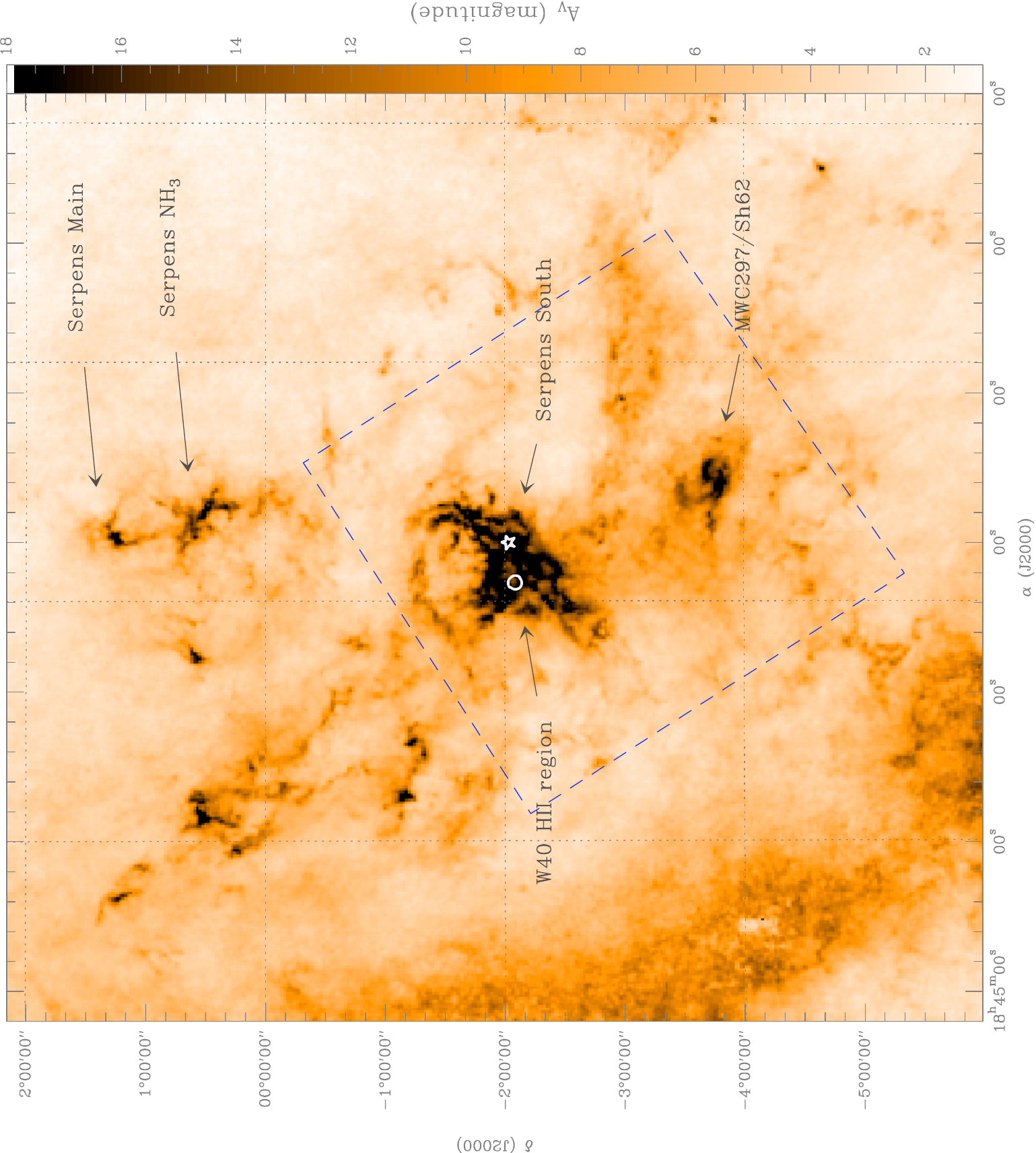}
      \caption{Visual extinction map towards the whole Aquila Rift / Serpens region derived by us from 2MASS data (see details in \citealp{schneider2010}). The spatial resolution is 2\amin ~FWHM. The dashed blue rectangle indicates the {\it Herschel} coverage. It comprises the bright H{\tiny II} region W40 (white circle), the Serpens South cluster (white star), and the H{\tiny II} region Sh2-62, associated with the young star MWC297. The Aquila Rift corresponds to the large elongated structure from the northeast to the southwest. Emisson from the Galactic plane is seen in the southeast corner. }
         \label{fig:overview}
   \end{figure}

\section{Observations}
\label{sect:obs}

The observations were performed in the parallel mode of  {\it Herschel} with a scanning speed of
60$''$/sec, which allows photometric imaging with SPIRE at 250, 350, and 500\,$\mu$m and PACS at 70 and 160\,$\mu$m. Two cross-linked scan maps were performed for a final coverage of $\sim 3.3^\circ \times 3.3^\circ$ (see Fig.~\ref{fig:overview}).

The SPIRE data were reduced using HIPE version 2.0 and modified
pipeline scripts; see \citet{griffin2010} for the in-orbit performance
and scientific capabilities, and \citet{swinyard2010} for calibration
methods and accuracy. A median baseline was applied to the maps for each scan leg, and the naive mapper was used as a mapmaking algorithm.  The PACS data
were reduced in HIPE 3.0. We used an updated version of the
calibration files following the most recent prescriptions of the PACS
ICC (see \citealp{konyves2010} for details). Multiresolution
median and second-order deglitching, as well as a high-pass filtering over the full scan leg length, were applied. The final
PACS maps were created using the photProject task, which performs
simple projection of the data cube on the map grid.

The resulting PACS maps are displayed in
Fig.~\ref{fig:pacs_images}. Owing to the rapid mapping speed, the
resulting point spread functions (PSFs) are elongated in the scan directions
leading to cross-like shapes of the PSFs with expected sizes of
5.9\asec$\times$12.2\asec ~at 70$\,\mu$m and 11.6\asec$\times$15.7\asec ~at 160$\,\mu$m. The resulting rms in these maps ranges from
50 to 1000 mJy/beam at 70$\,\mu$m and from 120 to 2200 mJy/beam at 160$\,\mu$m, depending on the level of background in the map. It is
in the W40/Sh2-64 H{\tiny II} region that the background level is the
highest.

\section{Overview and distance of the Aquila Rift complex}
\label{sect:overview}


The Aquila Rift is a coherent, $5^\circ$ long feature above the
Galactic plane at l=$28^\circ$, clearly visible on an extinction map
derived from the reddening of stars in 2MASS
(Fig.~\ref{fig:overview}).  A distance of 225$\,\pm\,$55$\,$pc has
been derived for this extinction wall, using spectro-photometric
studies of the optically visible stars \citep{straizys2003}.  This
distance is very similar to the usually adopted distance of 260$\,\pm\,$37$\,$pc for
the Serpens star-forming region\footnote{Note that a larger distance of 415$\,\pm\,$25$\,$pc has been recently claimed for Serpens Main based on a VLBA parallax of EC95, a young AeBe  star embedded in Serpens Main \citep{dzib2010}.}, located only $3^\circ$ north
\citep{straizys1996}. 

On the other hand, the most active and main extinction feature in the
2MASS extinction map is associated with the H{\small II} region
W40/Sh2-64, which has so far been considered to be at a distance
ranging from 100 and 700 pc depending on author (\citealp{smith1985,
vallee1987} and references therein). These distance estimates are
mostly based on kinematical distances that have large
uncertainties. W40 could therefore be at the same distance as
Serpens. Recently, \citet{gutermuth2008b} reported {\it Spitzer}
observations of an embedded cluster, referred to as Serpens South,
in the Aquila Rift region. This cluster is located very close in
projection on the sky to W40 (see Fig.~\ref{fig:overview}) and thus seems to be part of the W40
region. \citet{gutermuth2008b} proposed that the Serpens South cluster
should be part of Serpens since it has the same velocity ($6$
km/s). The molecular cloud associated with W40 and
traced by C{\small II} recombination lines and CO
\citep{zeilik&lada1978} has a velocity ranging from 4.5 to 6.5
km/s, which is also roughly the same as Serpens. More recent
N$_2$H$^+$ observations of the entire W40/Serpens South region confirm
similar velocities in the whole region with velocity differences of only
$\sim\,2\,$km/s (Maury et al. in prep). It is therefore more
straightforward to consider that the W40 region is 
a single complex at the same distance as Serpens. This distance also suits the MWC297 / Sh2-62 region since the young
10$\,$\msol ~star MWC297 itself has an accepted distance of 250 pc
\citep{drew1997}.
It is finally worth noting that the visual
extinction map by \citet{cambresy1999} derived from optical star
counts and only tracing the first layer of the
extinction wall 
has exactly the
same global aspect as the 2MASS extinction map of
Fig.~\ref{fig:overview}, suggesting that both Serpens Main and the W40 /
Aquila Rift / MWC297 region are associated with this extinction wall
at 260 pc. We thus adopt the distance of 260 pc for
the entire region in the following.

The 2MASS extinction map and the {\it Herschel} images (see Appendix for
PACS images and \citealp{konyves2010} for SPIRE images) clearly show a
massive cloud associated with W40. This cloud corresponds to
G28.74+3.52 in \citep{zeilik&lada1978} and has a mass of
$1.1\times10^4\,$\msol (derived from our 2MASS extinction map).  The
cloud associated with MWC297 is less massive ($4.1\times10^3\,$\msol),
and we obtain a total mass of 3.1$\times 10^4\,$\msol ~for the whole area covered by {\it Herschel}.



\onlfig{2}
{
 \begin{figure*}[hbtp]

 \includegraphics[angle=270,width=9.5cm]{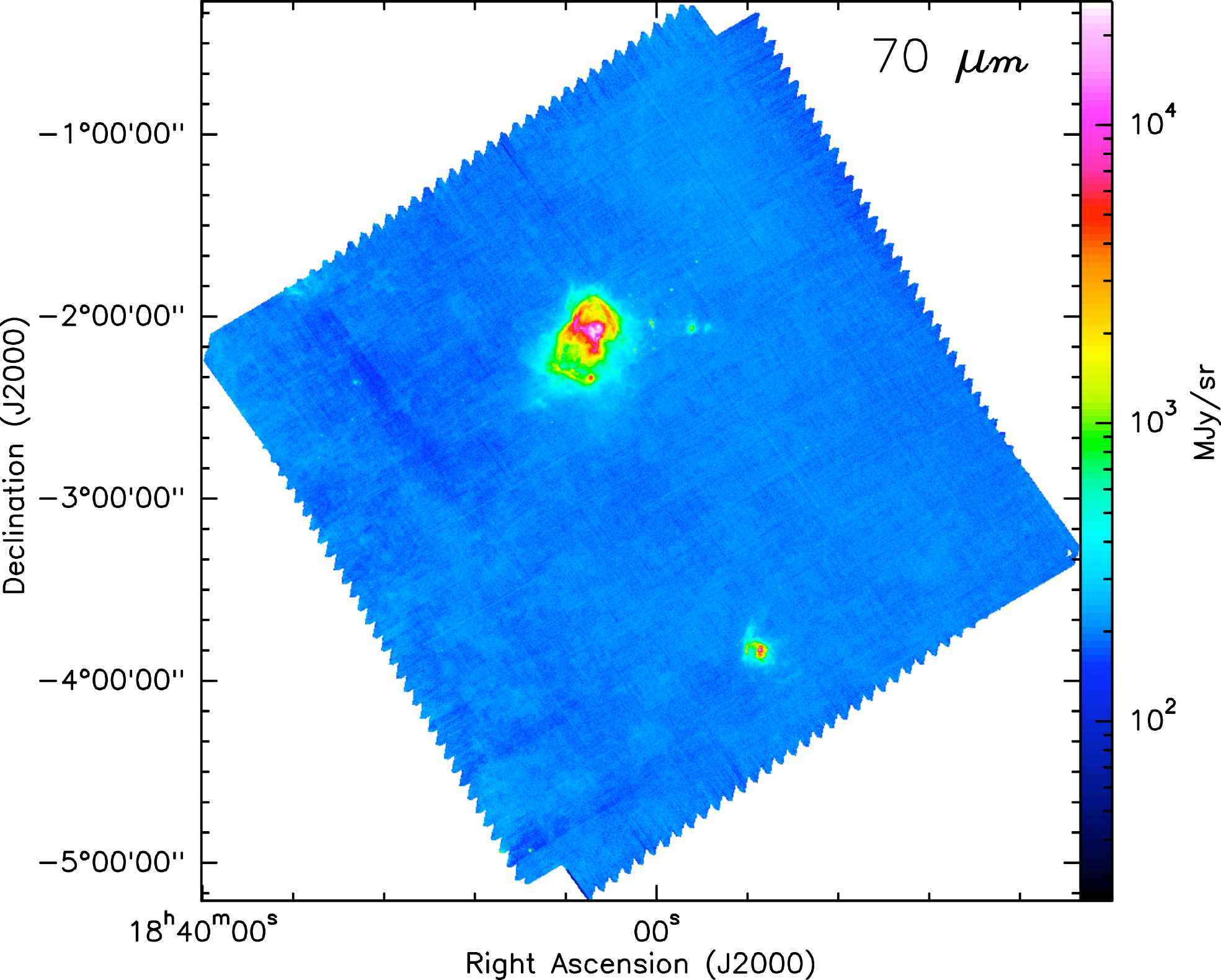}
 \includegraphics[angle=270,width=9.5cm]{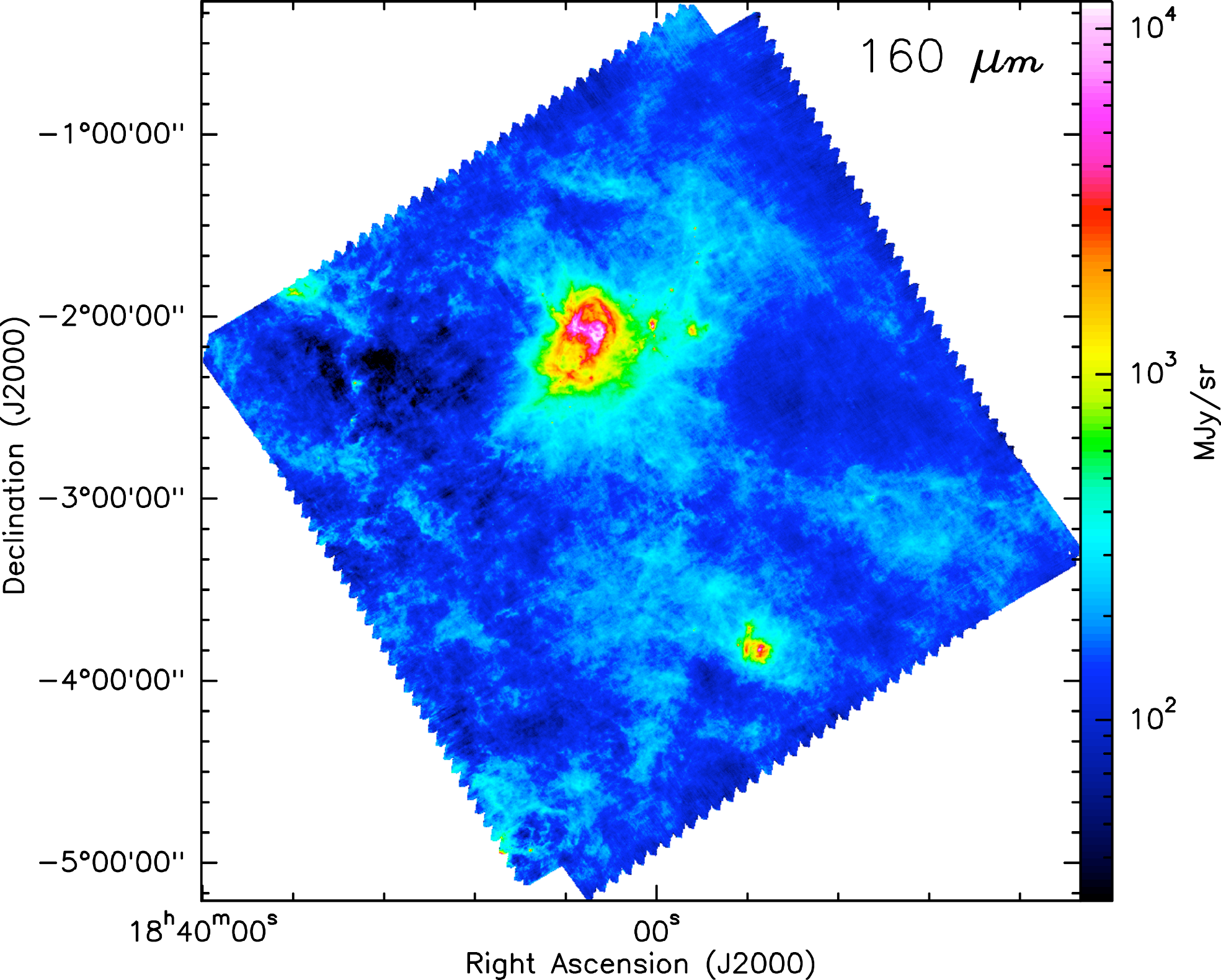}
 \mbox{}
  
 \caption{PACS 70$\,\mu$m (left) and 160$\,\mu$m (right) images of the Aquila field. See details about data reduction and map making in Sect.~\ref{sect:obs} and in \citet{konyves2010}. The corresponding SPIRE 250, 350, and 500$\,\mu$m images are shown in \citet{konyves2010}.}
         \label{fig:pacs_images}
   \end{figure*}
}

\onlfig{3}
{
 \begin{figure*}[hbtp]

 \includegraphics[angle=-90,width=16cm]{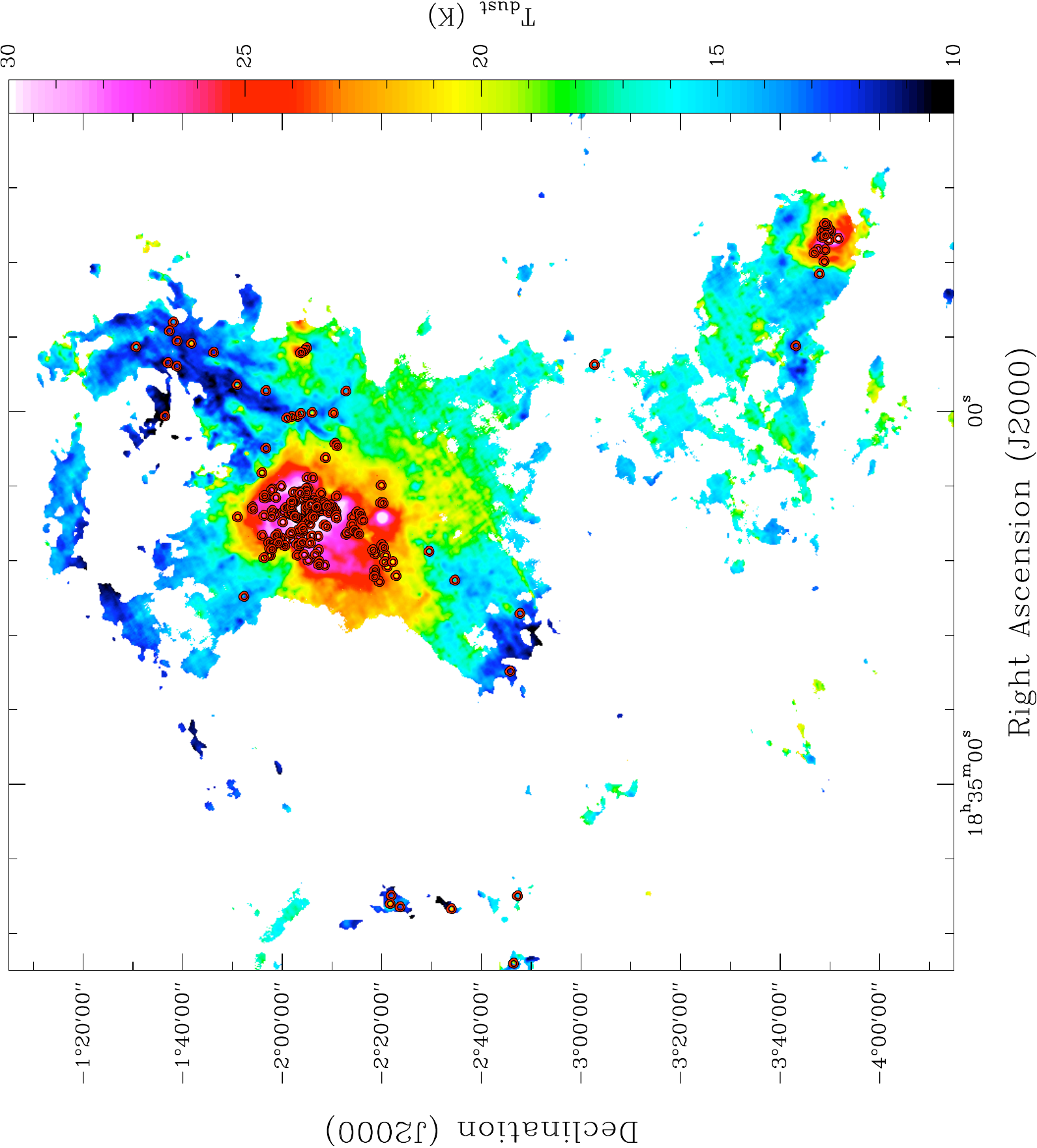}
 \mbox{}
  
 \caption{Distribution map of the 201 {\it Herschel} YSOs selected in Sect.~\ref{sect:sample}, over-plotted on the map of dust temperature. The dust temperature map was derived from graybody fits to the {\it Herschel} data (see details in \citealp{konyves2010}). }
         \label{fig:distribution}
   \end{figure*}
}

   \begin{figure*}[hbtp]
  \centering
  \includegraphics[angle=270,width=8.5cm]{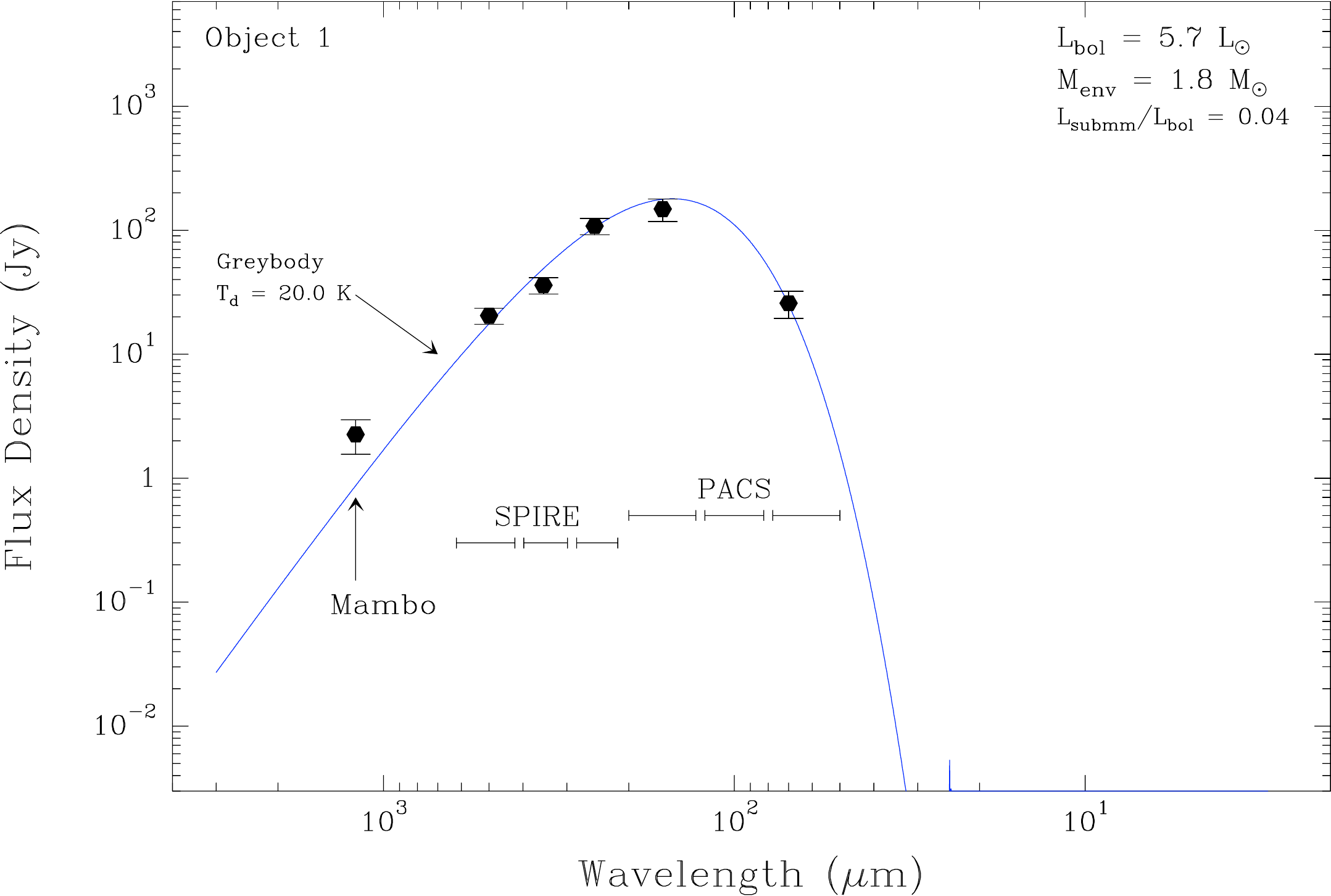}
  \includegraphics[angle=270,width=7.85cm]{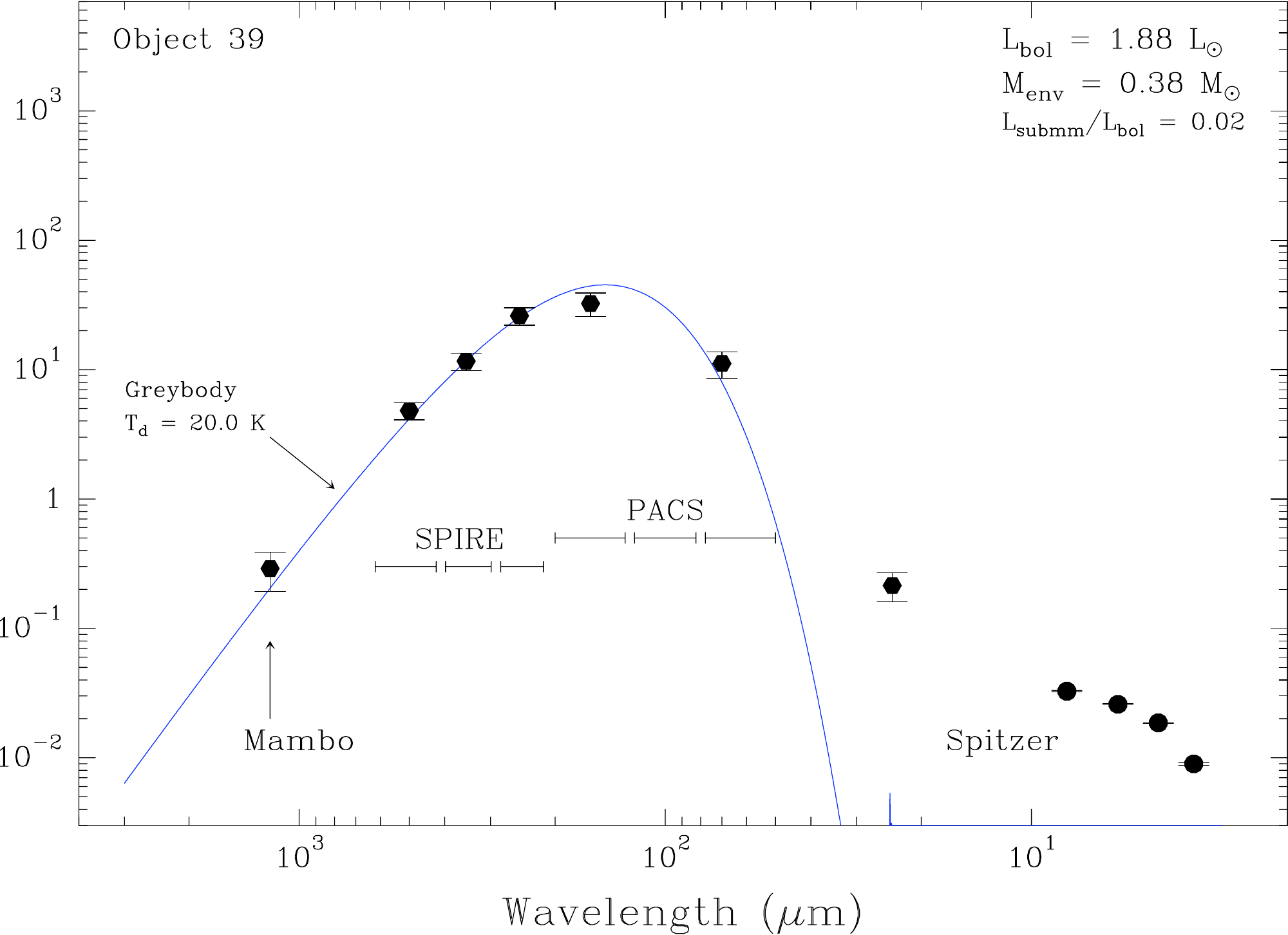}
      \caption{Spectral energy distributions of a newly discovered bright Class~0 (left panel) and of a weaker Class~0 object  (right panel), which was previously detected with {\it Spitzer} in \citet{gutermuth2008b}.}
         \label{fig:sed}
   \end{figure*}

\section{Results and analysis}
\label{sect:results}

\subsection{Source detection and identification of protostars}
\label{sect:sample}

A systematic source detection was performed on all 5 {\it
Herschel} bands using {\it getsources} \citep{menshchikov2010}. This code
uses a method based on a multiscale decomposition of the images to
disentangle the emission of a population of spatially coherent sources
in an optimized way in all bands simultaneously. 
We built a sample of the best candidate protostars for the
whole field. These sources are clearly detected in all {\it Herschel}
bands (high significance level), and we require a detection at
the shortest {\it Herschel} wavelength, 70$\,\mu$m (or 24$\,\mu$m when
{\it Spitzer} data were available), to distinguish YSOs from starless
cores.  Since the 24 and 70$\,\mu$m emission should only trace warm
dust from the inner regions of the YSO envelopes, these sources can be
safely interpreted as protostars. The 70$\,\mu$m fluxes have even been
recently recognized as a very good tracer of protostellar
luminosities \citep{dunham2008}. On the other hand, in the PDR region
of W40 some extended emission from warm dust at the H{\small II}
region interface could contaminate this YSO detection criterium. To avoid too stringent a contamination from this extended
emission, we selected only sources with an FWHM size smaller
than 40$^{\prime\prime}$ at 70$\,\mu$m.  Also, we had to make a source detection using a large pixel
size of 6\asec, which is good enough for starless cores mostly
detected in the SPIRE bands but not perfect to sample the
spatial resolution at 70$\,\mu$m and properly disentangle possible
multiple protostellar sources. A more precise detection could only be
achieved in a reduced area in the Serpens South region (see
Sect.~\ref{sect:closeup}).

A large number of compact sources are clearly seen in the 70$\,\mu$m
map down to the sensitivity limit of the survey. In the whole Aquila
field, 201 YSOs were detected with {\it getsources}. The best
achieved rms (50 mJy/beam) in the Aquila 70$\,\mu$m map in the lowest
background regions corresponds to a 5$\,\sigma$ detection
level in terms of protostar luminosity of 0.05$\,$\lsol ~using the
\citet{dunham2008} relationship.  In contrast, in the highest
background regions, the 5$\,\sigma$ detection level is then as high as
1.0$\,$\lsol. To account for the variable background level in Aquila,
we performed simulations to evaluate the final completeness level of
the YSO detection and obtained a 90~\% completeness level of
$\sim$0.2\,\lsol ~(see \citealp{konyves2010}), which is compatible with the above rough estimates using \citet{dunham2008}.

\subsection{Spatial distribution of the protostars }
\label{sect:distribution}

We plotted in Fig.~\ref{fig:distribution} the spatial
distribution of the {\it Herschel} sample of 201 YSOs overlaid on
the map of the dust temperature derived from a simple graybody fit of
the {\it Herschel} data (see details in \citealp{konyves2010}). It is clear that the W40 region
corresponds to the most active star-forming region in the {\it
Herschel} coverage with 90$\,$\% of the detected protostars. A second,
much less rich, site corresponds to MWC297 with 8$\,$\% of the
protostars, and another site to the east of W40 can be tentatively identified with
very few candidate protostars.


\subsection{Basic properties of the protostars: $M_{\rm env}$ and $L_{\rm bol}$}
\label{sect:sed}

For each source an SED was built
using the 5 bands of {\it Herschel}, as well as {\it Spitzer}
photometry \citep{gutermuth2008b} and MAMBO 1.2mm data (Maury et
al. in prep) when available. These SEDs were systematically fitted
using graybody functions to derive $M_{\rm env}$ in a systematic way, while the basic properties 
$L_{\rm bol}$, $L^{\rm \lambda > 350}_{\rm submm}$, and $T_{\rm bol}$ were obtained by simple integrations of the SEDs. Two
representative SEDs are displayed in Fig.~\ref{fig:sed} with a newly
discovered Class~0 object and a weaker Class 0 source, which has a {\it
Spitzer} counterpart in \citet{gutermuth2008b}.

   \begin{figure}[hbtp]
  \centering
  \includegraphics[angle=270,width=6cm]{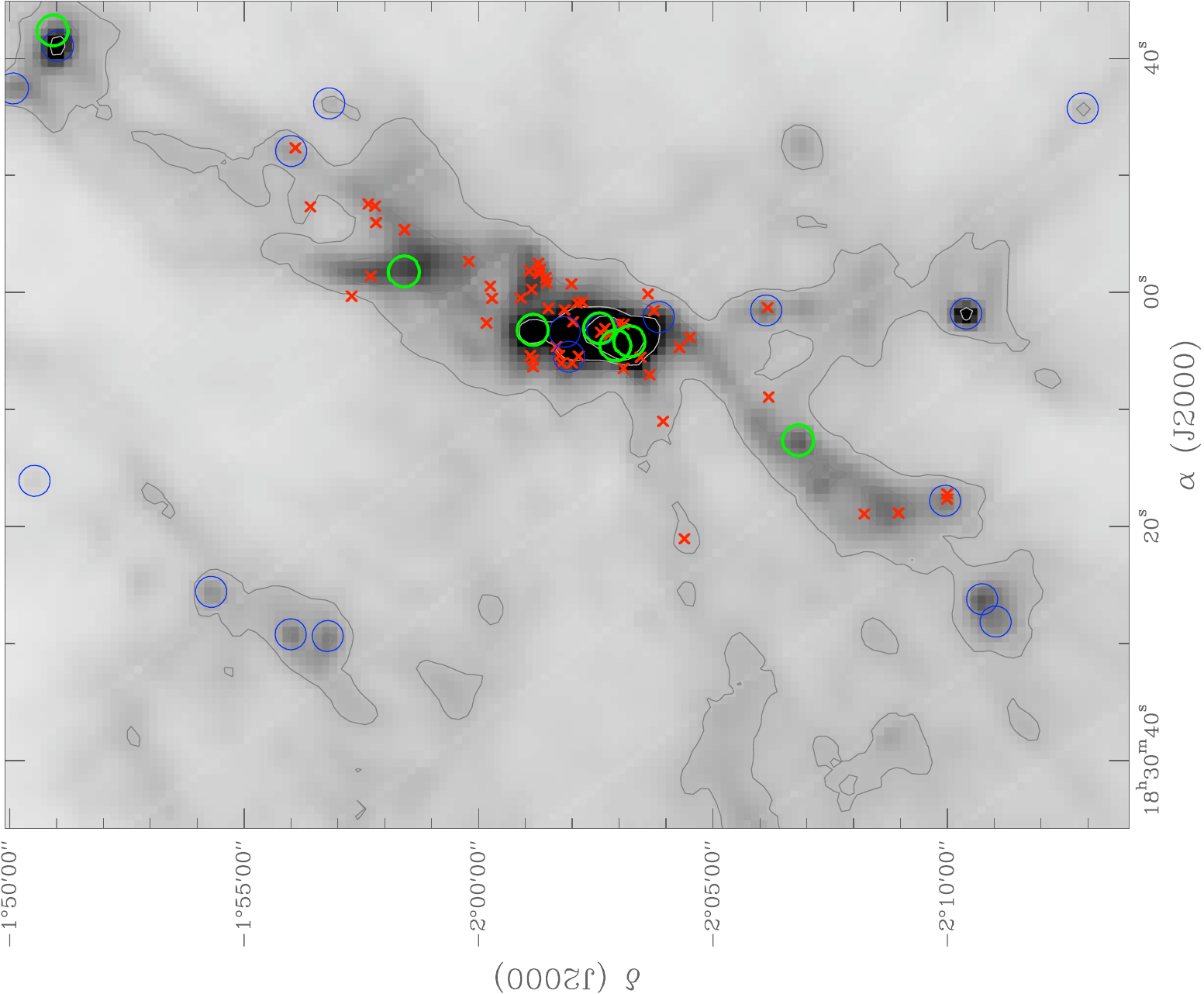}
      \caption{{\it Herschel} SPIRE 350$\mu$m image of the Serpens South region with the distribution of {\it Herschel} candidate protostars (blue circles) from the whole field extraction, of the 7 newly discovered Class~0 protostars (green circles), and of the {\it Spitzer} YSOs (red crosses; \citealp{gutermuth2008b}).}
         \label{fig:closeup}
   \end{figure}

\subsection{A close-up view of the Serpens South region}
\label{sect:closeup}

To go one step further in the identification and characterization of the {\it Herschel} protostars, we performed a more detailed analysis of the sources in a small area around the Serpens South cluster. In this area, we made a dedicated {\it getsources} source extraction using a smaller size pixel of 3\asec, and we could compare these first results with the {\it Spitzer} protostar population by \citet{gutermuth2008b}. 
We used {\it getsources} on 8 bands from 8 to 1200$\,\mu$m by adding the 8 and 24$\,\mu$m {\it Spitzer} and the 1.2mm MAMBO data to the 5 {\it Herschel} bands. 

A synthesized view of these first results based on this novel
panchromatic analysis of infrared to millimeter range data for this
area is given in Fig.~\ref{fig:closeup}. It shows the distribution of
{\it Herschel} protostars compared to the
{\it Spitzer} sources. The first analysis of this field
indicates that even in a highly clustered region like Serpens South, a
significant population of protostars were found to be missing by pure
near and mid-infrared imaging with as many as 7 newly detected Class~0 objects in this field. We also note that the {\it Spitzer}
protostars (most of these not detected with {\it Herschel}) probably
correspond to evolved or low-luminosity Class~I objects.

\section{Global view of the protostellar population in Aquila}
\label{sect:evol}

Using the basic properties derived in Sect.~\ref{sect:sed}, we can draw the first picture of the property
space {\it Herschel} is going to cover thanks to its
unprecedentedly sensitive and high spatial resolution in the
far-infrared.

\begin{figure}[hbtp]
%

 \includegraphics[angle=0,width=9.0cm]{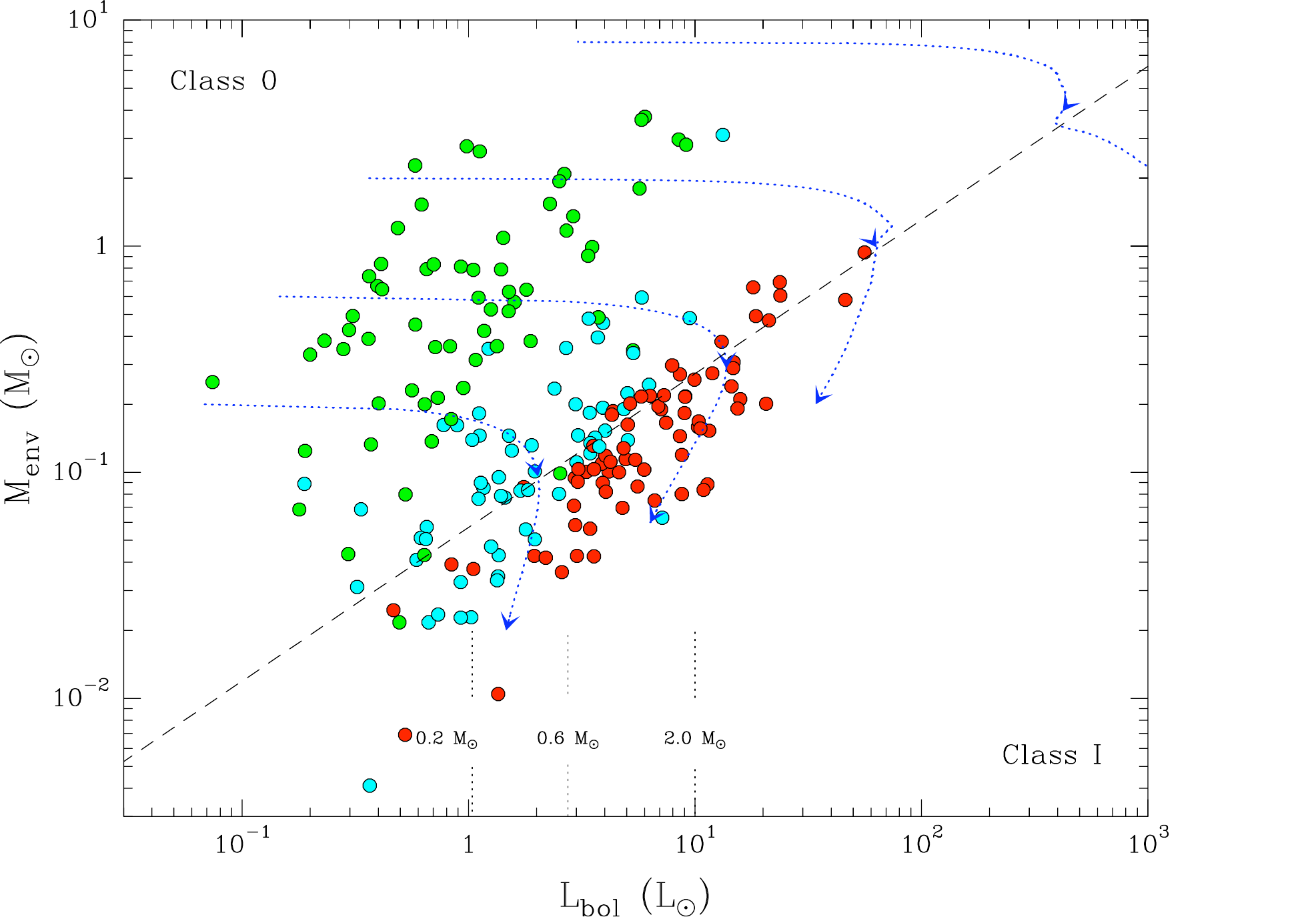}
  
      \caption{Distribution of the Herschel sample of protostars in the protostellar evolutionary diagram $M_{\rm env} - L_{\rm bol}$.  The green and red circles are for the Class~0 and Class~I protostars with $L^{\rm \lambda > 350}_{\rm submm}/L^{\rm 70- 500}_{\rm bol} > 0.03$ and $< 0.01$ (see text), respectively. The intermediate, more uncertain cases with  $0.03 > L^{\rm \lambda > 350}_{\rm submm}/L^{\rm 70- 500}_{\rm bol} > 0.01$ are displayed as light blue circles. The evolutionary tracks for 0.2, 0.6, 2.0, and 8.0$\,$\msol ~are displayed as dotted curves. The formal separation between Classes~0 and I in this diagram corresponds to 50~\% of the mass accreted, which corresponds to the locii of the first arrows on the curves (see also the dashed separating line). The second arrows on the curves indicate 90~\% of the mass accreted. }
         \label{fig:evol}
   \end{figure}

In Fig.~\ref{fig:evol} we plotted the location of
the 201 {\it Herschel} YSOs obtained in the entire field in a
$M_{\rm env} - L_{\rm bol}$ evolutionary diagram used to
compare observed properties with theoretical evolutionary models or
tracks.  The displayed tracks represent the expected evolution of
protostars of masses 0.2, 0.6, 2.0, and 8.0$\,$\msol ~from the
earliest times of accretion (upper left part of the diagram) to the
time of 50~\% mass accreted (conceptual limit between Class~0 and
Class~I YSOs), and the time for 90~\% mass accreted (see
\citealp{bontemps1996, saraceno1996, andre2000, andre2008}). In this plot, we distinguished objects with an  $L^{\rm \lambda > 350}_{\rm submm}/L^{\rm 70- 500}_{\rm bol}$ higher than 0.03 which could be safely recognized as Class~0 objects, from YSOs with $L^{\rm \lambda > 350}_{\rm submm}/L^{\rm 70- 500}_{\rm bol}$ lower than 0.01 which are proposed to be Class~I sources. The intermediate objects with $0.03>L^{\rm \lambda > 350}_{\rm submm}/L^{\rm 70- 500}_{\rm bol}>0.01$ should be seen as objects with an uncertain classification. A forthcoming analysis will resolve their nature by building complete SEDs including Spitzer data for a large part of the Aquila field. So far we could safely classify objects only in the reduced area of Serpens South (Sect.~\ref{sect:closeup}).  In this subfield, we verified that objects with $L^{\rm \lambda > 350}_{\rm submm}/L^{\rm 70- 500}_{\rm bol} > 0.03$ and $T^{\rm 70- 500}_{\rm bol} < 27\,$K using the reduced  (only the 5 {\it Herschel} bands from 70 to 500$\,\mu$m) SED coverage are indeed all found to be Class~0 objects based on the full coverage from 8$\,\mu$m to 1.2$\,$mm. We see that the obtained location of Class~0 and Class~I YSOs is compatible with the 50~\% mass accreted limit. Imposing Class~0 objects to have to be above this limit (dashed line in Fig.~\ref{fig:evol}), we finally found between 45 (for $T^{\rm 70- 500}_{\rm bol} < 27\,$K) and 60 ($L^{\rm \lambda > 350}_{\rm submm}/L^{\rm 70- 500}_{\rm bol} > 0.03$) Class~0 objects in the entire field of Aquila.

In conclusion, even if
the precise locations of the {\it Herschel} protostars in this diagram
are seen as a preliminary result and will be updated with a more
complete analysis and source detection, our early results clearly indicate that {\it Herschel} is a powerful tool for probing the virtually
unexplored area of the physical properties of the earliest stages of protostellar evolution.

%

%

\begin{acknowledgements}
 SPIRE has been developed by a consortium of institutes led by
Cardiff Univ. (UK) and including Univ. Lethbridge (Canada);
NAOC (China); CEA, LAM (France); IFSI, Univ. Padua (Italy);
IAC (Spain); Stockholm Observatory (Sweden); Imperial College
London, RAL, UCL-MSSL, UKATC, Univ. Sussex (UK); Caltech, JPL,
NHSC, Univ. Colorado (USA). This development has been supported
by national funding agencies: CSA (Canada); NAOC (China); CEA,
CNES, CNRS (France); ASI (Italy); MCINN (Spain); SNSB (Sweden);
STFC (UK); and NASA (USA).
PACS has been developed by a consortium of institutes led by MPE (Germany) and including UVIE (Austria); KUL, CSL, IMEC (Belgium); CEA, LAM (France); MPIA (Germany); IFSI, OAP/AOT, OAA/CAISMI, LENS, SISSA (Italy); IAC (Spain). This development has been supported by the funding agencies BMVIT (Austria), ESA-PRODEX (Belgium), CEA/CNES (France), DLR (Germany), ASI (Italy), and CICT/MCT (Spain). We thanks Rob Gutermuth for providing us with the list of Spitzer sources in the Serpens South sub-field prior to publication.

\end{acknowledgements}

\bibliography{biblio}

\begin{thebibliography}{28}
\expandafter\ifx\csname natexlab\endcsname\relax\def\natexlab#1{#1}\fi

\bibitem[{{Allen} {et~al.}(2007){Allen}, {Megeath}, {Gutermuth}, {Myers},
  {Wolk}, {Adams}, {Muzerolle}, {Young}, \& {Pipher}}]{allen2007}
{Allen}, L., {Megeath}, S.~T., {Gutermuth}, R., {et~al.} 2007, Protostars and
  Planets V, 361

\bibitem[{{Andr{\'e}} {et~al.}(2010){Andr{\'e}}, {Men'shchikov}, {Bontemps},
  {et~al.}}]{andre2010}
{Andr{\'e}}, P., {Men'shchikov}, A., {Bontemps}, S., {et~al.} 2010, \aap, this
  volume

\bibitem[{{Andr{\'e}} {et~al.}(2008){Andr{\'e}}, {Minier}, {Gallais},
  {Rev{\'e}ret}, {Le Pennec}, {Rodriguez}, {Boulade}, {Doumayrou}, {Dubreuil},
  {Lortholary}, {Martignac}, {Talvard}, {De Breuck}, {Hamon}, {Schneider},
  {Bontemps}, {Lagage}, {Pantin}, {Roussel}, {Miller}, {Purcell}, {Hill}, \&
  {Stutzki}}]{andre2008}
{Andr{\'e}}, P., {Minier}, V., {Gallais}, P., {et~al.} 2008, \aap, 490, L27

\bibitem[{{Andr\'e} {et~al.}(2000){Andr\'e}, {Ward-Thompson}, \&
  {Barsony}}]{andre2000}
{Andr\'e}, P., {Ward-Thompson}, D., \& {Barsony}, M. 2000, Protostars and
  Planets IV, 59

\bibitem[{{Bontemps} {et~al.}(2001){Bontemps}, {Andr{\'e}}, {Kaas}, {Nordh},
  {Olofsson}, {Huldtgren}, {Abergel}, {Blommaert}, {Boulanger}, {Burgdorf},
  {Cesarsky}, {Cesarsky}, {Copet}, {Davies}, {Falgarone}, {Lagache},
  {Montmerle}, {P{\'e}rault}, {Persi}, {Prusti}, {Puget}, \&
  {Sibille}}]{bontemps2001}
{Bontemps}, S., {Andr{\'e}}, P., {Kaas}, A.~A., {et~al.} 2001, \aap, 372, 173

\bibitem[{{Bontemps} {et~al.}(1996){Bontemps}, {Andr\'e}, {Terebey}, \&
  {Cabrit}}]{bontemps1996}
{Bontemps}, S., {Andr\'e}, P., {Terebey}, S., \& {Cabrit}, S. 1996, \aap, 311,
  858

\bibitem[{{Cambr{\'e}sy}(1999)}]{cambresy1999}
{Cambr{\'e}sy}, L. 1999, \aap, 345, 965

\bibitem[{{Di Francesco} {et~al.}(2007){Di Francesco}, {Evans}, {Caselli},
  {Myers}, {Shirley}, {Aikawa}, \& {Tafalla}}]{difrancesco2007}
{Di Francesco}, J., {Evans}, II, N.~J., {Caselli}, P., {et~al.} 2007, in
  Protostars and Planets V, ed. B.~{Reipurth}, D.~{Jewitt}, \& K.~{Keil},
  17--32

\bibitem[{{Drew} {et~al.}(1997){Drew}, {Busfield}, {Hoare}, {Murdoch}, {Nixon},
  \& {Oudmaijer}}]{drew1997}
{Drew}, J.~E., {Busfield}, G., {Hoare}, M.~G., {et~al.} 1997, \mnras, 286, 538

\bibitem[{{Dunham} {et~al.}(2008){Dunham}, {Crapsi}, {Evans}, {Bourke},
  {Huard}, {Myers}, \& {Kauffmann}}]{dunham2008}
{Dunham}, M.~M., {Crapsi}, A., {Evans}, II, N.~J., {et~al.} 2008, \apjs, 179,
  249

\bibitem[{{Dzib} {et~al.}(2010){Dzib}, {Loinard}, {Mioduszewski}, {Boden},
  {Rodriguez}, \& {Torres}}]{dzib2010}
{Dzib}, S., {Loinard}, L., {Mioduszewski}, A.~J., {et~al.} 2010,
  ArXiv:1003.5900

\bibitem[{{Evans} {et~al.}(2009){Evans}, {Dunham}, {J{\o}rgensen}, {Enoch},
  {Mer{\'{\i}}n}, {van Dishoeck}, {Alcal{\'a}}, {Myers}, {Stapelfeldt},
  {Huard}, {Allen}, {Harvey}, {van Kempen}, {Blake}, {Koerner}, {Mundy},
  {Padgett}, \& {Sargent}}]{evans2009}
{Evans}, N.~J., {Dunham}, M.~M., {J{\o}rgensen}, J.~K., {et~al.} 2009, \apjs,
  181, 321

\bibitem[{{Griffin} {et~al.}(2010)}]{griffin2010}
{Griffin}, M. {et~al.} 2010, \aap, this volume

\bibitem[{{Gutermuth} {et~al.}(2008){Gutermuth}, {Bourke}, {Allen}, {Myers},
  {Megeath}, {Matthews}, {J{\o}rgensen}, {Di Francesco}, {Ward-Thompson},
  {Huard}, {Brooke}, {Dunham}, {Cieza}, {Harvey}, \&
  {Chapman}}]{gutermuth2008b}
{Gutermuth}, R.~A., {Bourke}, T.~L., {Allen}, L.~E., {et~al.} 2008, \apjl, 673,
  L151

\bibitem[{{Kaas} {et~al.}(2004){Kaas}, {Olofsson}, {Bontemps}, {Andr{\'e}},
  {Nordh}, {Huldtgren}, {Prusti}, {Persi}, {Delgado}, {Motte}, {Abergel},
  {Boulanger}, {Burgdorf}, {Casali}, {Cesarsky}, {Davies}, {Falgarone},
  {Montmerle}, {Perault}, {Puget}, \& {Sibille}}]{kaas2004}
{Kaas}, A.~A., {Olofsson}, G., {Bontemps}, S., {et~al.} 2004, \aap, 421, 623

\bibitem[{{K\"onyves} {et~al.}(2010){K\"onyves}, {Andr{\'e}}, {Men'shchikov},
  {et~al.}}]{konyves2010}
{K\"onyves}, V., {Andr{\'e}}, P., {Men'shchikov}, A., {et~al.} 2010, \aap, this
  volume

\bibitem[{{Men'shchikov} {et~al.}(2010){Men'shchikov}, {Andr{\'e}}, {Didelon},
  {et~al.}}]{menshchikov2010}
{Men'shchikov}, A., {Andr{\'e}}, P., {Didelon}, P., {et~al.} 2010, \aap, this
  volume

\bibitem[{{Nordh} {et~al.}(1996){Nordh}, {Olofsson}, {Abergel}, {Andre},
  {Blommaert}, {Bontemps}, {Boulanger}, {Burgdorf}, {Cesarsky}, {Cesarsky},
  {Copet}, {Davies}, {Falgarone}, {Huldtgren}, {Kaas}, {Lagache}, {Montmerle},
  {Perault}, {Persi}, {Prusti}, {Puget}, \& {Sibille}}]{nordh1996}
{Nordh}, L., {Olofsson}, G., {Abergel}, A., {et~al.} 1996, \aap, 315, L185

\bibitem[{{Pilbratt} {et~al.}(2010)}]{pilbratt2010}
{Pilbratt}, G. {et~al.} 2010, \aap, this volume

\bibitem[{{Poglitsch} {et~al.}(2010)}]{poglitsch2010}
{Poglitsch}, G. {et~al.} 2010, \aap, this volume

\bibitem[{{Saraceno} {et~al.}(1996){Saraceno}, {Andr\'e}, {Ceccarelli},
  {Griffin}, \& {Molinari}}]{saraceno1996}
{Saraceno}, P., {Andr\'e}, P., {Ceccarelli}, C., {Griffin}, M., \& {Molinari},
  S. 1996, \aap, 309, 827

\bibitem[{{Schneider} {et~al.}(2010){Schneider}, {Bontemps}, {Simon},
  {Ossenkopf}, {Federrath}, {Klessen}, {Motte}, \& {Brunt}}]{schneider2010}
{Schneider}, N., {Bontemps}, S., {Simon}, R., {et~al.} 2010, ArXiv:1001.2453

\bibitem[{{Smith} {et~al.}(1985){Smith}, {Bentley}, {Castelaz}, {Gehrz},
  {Grasdalen}, \& {Hackwell}}]{smith1985}
{Smith}, J., {Bentley}, A., {Castelaz}, M., {et~al.} 1985, \apj, 291, 571

\bibitem[{{Strai{\v z}ys} {et~al.}(1996){Strai{\v z}ys}, {{\v C}ernis}, \&
  {Barta{\v s}i{\= u}t{\.e}}}]{straizys1996}
{Strai{\v z}ys}, V., {{\v C}ernis}, K., \& {Barta{\v s}i{\= u}t{\.e}}, S. 1996,
  Baltic Astronomy, 5, 125

\bibitem[{{Strai{\v z}ys} {et~al.}(2003){Strai{\v z}ys}, {{\v C}ernis}, \&
  {Barta{\v s}i{\= u}t{\.e}}}]{straizys2003}
{Strai{\v z}ys}, V., {{\v C}ernis}, K., \& {Barta{\v s}i{\= u}t{\.e}}, S. 2003,
  \aap, 405, 585

\bibitem[{{Swinyard} {et~al.}(2010){Swinyard}, {Ade}, {Baluteau},
  {et~al.}}]{swinyard2010}
{Swinyard}, B.~M., {Ade}, P., {Baluteau}, J.~P., {et~al.} 2010, \aap, this
  volume

\bibitem[{{Vallee}(1987)}]{vallee1987}
{Vallee}, J.~P. 1987, \aap, 178, 237

\bibitem[{{Zeilik} \& {Lada}(1978)}]{zeilik&lada1978}
{Zeilik}, II, M. \& {Lada}, C.~J. 1978, \apj, 222, 896

\end{thebibliography}
\bibliographystyle{aa}

\end{document}